\documentclass{iau307}
\usepackage{graphicx}
\usepackage{natbib}
\usepackage{url}
\usepackage{dtklogos}
\bibpunct{(}{)}{;}{a}{}{,}

\title[]
{Stellar magnetic fields from four Stokes parameter observations}

\author[Rusomarov et al.]   
{N. Rusomarov$^1$
 \and O. Kochukhov$^1$
 \and N. Piskunov$^1$}

\affiliation{$^1$Uppsala University \\ email: {\tt naum.rusomarov@physics.uu.se}}

\pubyear{2014}
\volume{307} 
\pagerange{}
\jname{New windows on massive stars: asteroseismology, interferometry, and spectropolarimetry}
\editors{G. Meynet, C. Georgy, J.H. Groh \& Ph. Stee, eds.}

\begin{document}

\maketitle

\begin{abstract}
Magnetic Doppler imaging from four Stokes parameter observations can uncover new information that is of interest to the evolution and structure of magnetic fields of intermediate and high-mass stars. 

Our MDI study of the chemically peculiar star HD\,24712 from four Stokes parameter observations, obtained with the HARPSpol instrument at the 3.6-m ESO telescope, revealed a magnetic field with strong dipolar component and weak small-scale contributions.

This finding gives evidence for the hypothesis that old Ap stars have predominantly dipolar magnetic fields. 
\end{abstract}

Magnetic Doppler imaging (MDI) from four Stokes parameter observations contributes to the understanding of the evolution and structure of magnetic fields of intermediate and high-mass main sequence stars. Such studies have recently revealed unprecedented previously unknown level of detail of stellar magnetic fields.

Along these lines, we have started a new program aimed at obtaining phase-resolved, spectropolarimetric observations of chemically peculiar stars in all four Stokes parameters, and using these data to perform MDI inversions for these stellar objects. Thus far, we have achieved full rotational phase coverage of the chemically peculiar stars HD\,24712, HD\,125248, and HD\,119419. The observational data for these objects were obtained with the HARPSpol instrument at the 3.6-m ESO telescope at La Silla, Chile. This spectrograph allows full Stokes vector observations with a spectral resolution greater than $10^5 $. An analysis of the full Stokes vector spectropolarimetric data set of HD\,24712 has been published by \citet{Rusomarov2013p8}.

\begin{figure}
  \centering
  \includegraphics[width=0.8\textwidth, height=0.35\textheight]{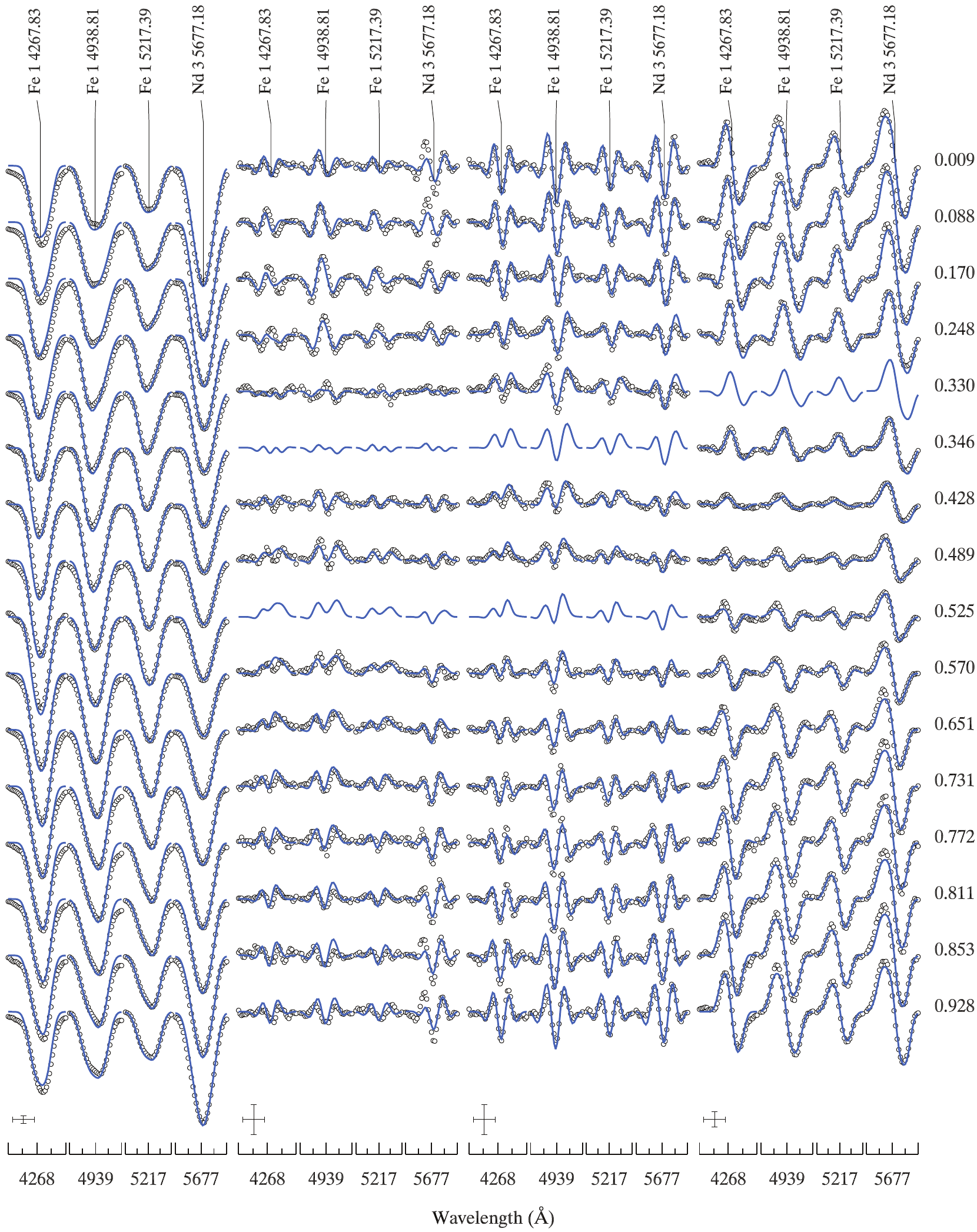}
  \caption{Comparison of the observed (symbols) and calculated (lines) Stokes profiles for HD\,24712. The bars at the lower left of each panel show the horizontal and vertical scale (0.2Å and 2.5\% of the Stokes~$I$ continuum intensity).}
  \label{fig:rusomarov_fig1}
\end{figure}

\begin{figure}
  \centering
  \includegraphics[scale=0.25]{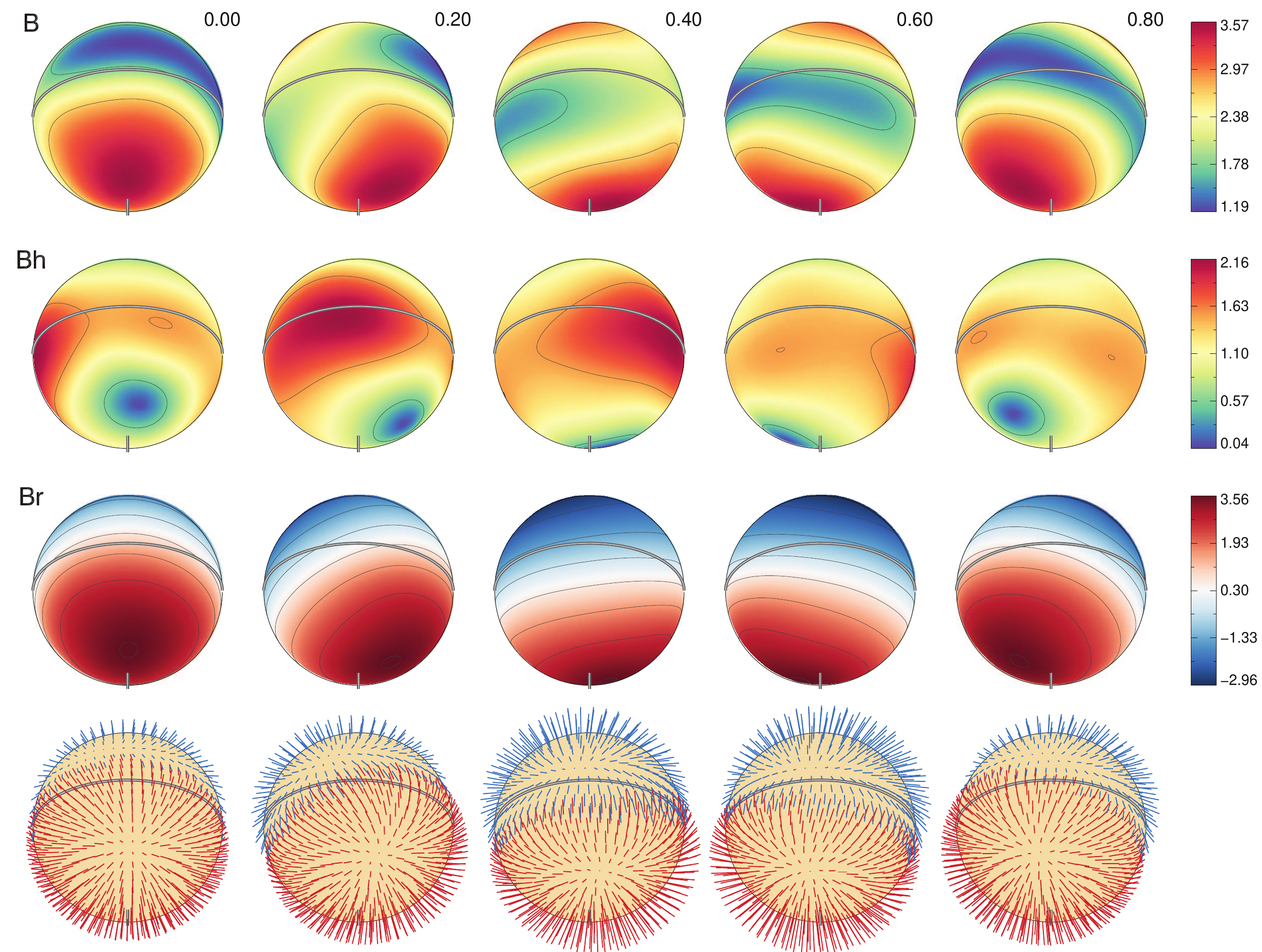}
  \caption{Distribution of the magnetic field on the surface of HD\,24712. Magnetic field modulus (first row), horizontal component (second row), radial field component (third row), and field orientation (last row). The color bars indicate field strength in kG. The arrows length is proportional to the field strength. The contours are plotted with 1\,kG step.}
  \label{fig:rusomarov_fig2}
\end{figure}

Here, we present the results of the MDI analysis of HD\,24712. We derived chemical abundance and magnetic distribution maps from a selection of Fe, Nd, and Na lines. The comparison of observed and calculated Stokes profiles of several spectral lines is shown in Fig.~\ref{fig:rusomarov_fig1}. The spherical projections of the magnetic maps are shown in Fig.~\ref{fig:rusomarov_fig2}.

Our analysis shows that the magnetic field topology of HD\,24712 is mostly poloidal and dipolar with small contributions from higher-order harmonics. In contrast, the only two other stars for which we have MDI studies from four Stokes parameter observations, $\alpha^2$\,CVn \citep{Kochukhov10p13,Silvester2014p182} and 53\,Cam \citep{Kochukhov2004p613}, exhibit dipole-like fields with considerably more complexity on smaller scales.

Numerical MHD simulations of fossil fields in main sequence stars with radiative envelopes \citep{Braithwaite06p1077} showed that stable magnetic field configurations can exist in the form of a poloidal magnetic field wrapped around toroidal flux tubes. The authors also proposed that stars possessing dipole-like fields with little small-scale structures are older than stars with more complex fields.

The finding of dipole, axisymmetric field for HD\,24712, which is the oldest object in our three stars sample, in conjunction with the results for $\alpha^2$\,CVn and 53\,Cam, supports this hypothesis. In the future, we hope to perform MDI of other Ap/Bp stars with different masses and ages, which will help us further assess this hypothesis and investigate the dependence of the magnetic field geometry on other stellar parameters.

\bibliographystyle{iau307}
\bibliography{rusomarov}

\begin{thebibliography}{}

\bibitem[\protect\astroncite{Braithwaite \&
  Nordlund}{2006}]{Braithwaite06p1077}
Braithwaite, J. \& Nordlund, {\AA}. 2006,
\newblock {\em \aap} 450, 1077

\bibitem[\protect\astroncite{{Kochukhov} et~al.}{2004}]{Kochukhov2004p613}
{Kochukhov}, O., {Bagnulo}, S., {Wade}, G.~A., {et~al.} 2004,
\newblock {\em \aap} 414, 613

\bibitem[\protect\astroncite{Kochukhov \& Wade}{2010}]{Kochukhov10p13}
Kochukhov, O. \& Wade, G.~A. 2010,
\newblock {\em \aap} 513, A13

\bibitem[\protect\astroncite{Rusomarov et~al.}{2013}]{Rusomarov2013p8}
Rusomarov, N., Kochukhov, O., Piskunov, N., {et~al.} 2013,
\newblock {\em \aap} 558, A8

\bibitem[\protect\astroncite{{Silvester} et~al.}{2014}]{Silvester2014p182}
{Silvester}, J., {Kochukhov}, O., \& {Wade}, G.~A. 2014,
\newblock {\em \mnras} 440, 182

\end{thebibliography}

\end{document}